**Modification of Newtonian Gravity: Implications for Hot Gas in Clusters and Galactic Angular Momentum**

**Louise Rebecca [a], Dominic Sebastian [a], C Sivaram [b], Kenath Arun *[a]**

[a] Department of Physics and Electronics, CHRIST (Deemed to be University), Bangalore, India

[b] Indian Institute of Astrophysics, Bangalore, India

**Abstract:** In view of the negative results from various dark matter detection experiments, we had earlier proposed an alternate theoretical framework through Modification of Newtonian Gravity (MONG). Here, the Poison's equation is modified by introducing an additional gravitational self-energy density term along with the usual dark energy density term. In this work we extend this model to account for the presence of low-density gas at high temperatures $(10^8 K)$ in the intra cluster medium (ICM) by estimating the velocities to which particles will be subjected by the modified gravitational force. Considering that the ICM is under the influence of the cluster's gravity, particle velocities of the ions in the ICM must be balanced by the cluster's gravitational force. The particle velocities obtained for various clusters from their temperature profiles match the velocity produced by the MONG gravitational force. Thus, the increase in the gravitational potential at the outskirts of galaxies balances the thermal pressure of the ICM, maintaining hydrostatic equilibrium without invoking DM. The effect of MONG on the angular momentum of galaxies is also studied by obtaining a scaling relation between the angular momentum and the mass of a galaxy. MONG predicts a higher dependence on mass in comparison to the ΛCDM model. This increased dependence on mass compensates for the halo contribution to the angular momentum. The angular momentum from MONG for galaxies from the SPARC database is compared to the halo angular momentum by a Chi-square fit technique. The correlation coefficient is found to be unity, showing a replicable result.

*Corresponding author:

Kenath Arun: e-mail: kenath.arun@christuniversity.in; https://orcid.org/0000-0002-2183-9425

## 1. Introduction

The observations of the big bang nucleosynthesis and cosmic microwave background radiation suggest that the universe is made up of 4.9% ordinary baryonic matter, 26.8% dark matter, and 68.3% dark energy [1]. Other compelling evidence such as the dynamics of large-scale structures such as galaxy and galaxy clusters, gravitational lensing, the presence of hot gas between galaxies, etc., conclude that dark matter makes up approximately 85% of the matter in the Universe.

With the wide range of observational evidence pointing to the need for dark matter, various dark matter detection experiments have been running for the last couple of decades [2,3]. Direct detection experiments such as the LZ, XENON, XENONIT, and PANDA [4,5] look for signatures from the interaction between Weakly Interacting Massive Particles (WIMPs) (a candidate for DM) and baryonic matter [6,7]. The most recent upgrade to XENON1T is the XENONnT (2020-present), with a total xenon mass of more than eight tons. So far, the detector has registered only 16 events, most of which is due to electronic recoils or neutron collisions, and has only limited the interaction cross-section to $10^{-48} cm^2$ with no conclusive evidence of dark matter [8,9].

Though the observational evidence for dark matter is quite compelling, it remains undetected by various DM detection experiments. Also, there are several candidates (about 60) for DM particles ranging from axions (one millionth the electron mass) to wimpzillas (trillion times the proton mass). Several other candidates resulting from particle physics are supersymmetric particles like gluinos, and gravitinos, apart from q-balls, quark nuggets, etc. It is unclear whether DM is only one of these particles or a combination of these candidates and, if so, in what proportions. All these uncertainties imply the need to seriously consider alternate models to DM that could explain the cosmological phenomena that are usually attributed to DM [10].

One such alternate model is the Modification of Newtonian Dynamics (MOND) in which Newtonian law gets modified for acceleration below a fundamental acceleration, $a_0 \approx 10^{-8} cms^{-2}$ [11,12]. MOND is successful in explaining galactic dynamics, but involves the ad hoc introduction of $a_0$. MOND is a nonrelativistic theory that works best in accounting for the dynamics of galaxies and galaxy clusters [13]. There are other non-local models that have addressed this issue, see for example [14]. Though MOND works well in galactic scales, it fails to produce relativistic effects needed to explain CMB fluctuations and gravitational lensing effects. In alternate gravity theories (including MONG), the absence of DM could lead to the decrease in the amplitude in each of the subsequent peaks in the matter power spectrum as baryonic damping would dominate. CDM particles on the other hand provide additional force on the oscillations, thereby producing the observed third peak. Though the amplitude ratio is not a priori prediction in ΛCDM, its flexibility provides a posteriori fit to the data. This is not possible in MOND. Though MOND lacks a parent relativistic model, a scalar field of the sort invoked in TeVeS is shown to have the same effect on the oscillations as ΛCDM. Thus, the third peak would help in arriving at a complete MOND theory [15].

TeVeS, a relativistic MOND theory based on the action principle, involves a scalar and 4-vector field with a free function. It predicts gravitational lensing without dark matter, passes solar tests of GR, and avoids superluminal propagation [16]. While the theory makes no predictions for MOND effects on CMB, a new relativistic theory (RMOND) resembling TeVeS but with extra degrees of freedom, aligns with observed CMB and matter spectra (MPS) on linear cosmological scales. Relativistic MOND theories that make precise predictions of gravitational lensing and cosmology have been proposed earlier [17,18,19,20]. No theory so far has predicted CMB and matter power spectrum (MPS) while retaining the MOND theory for galactic dynamics. The relativistic MOND (RMOND) theory reproduces lensing and galactic predictions similar to TeVeS, while also reproducing CMB and MPS [21]. A recent work demonstrated that variations in symmetries of General Relativity at large scales produce effects similar to dark matter [22].

## 2. Modification of Newtonian Gravity (MONG)

An alternate explanation for flat rotation curves of galaxies can be arrived at through Modifications of Newtonian Gravity (MONG). In MONG, the Poisson's equation gets modified by adding a gravitational self-energy density term along with the dark energy cosmological constant term. In modified gravity theories, gravitational self-energy would contribute to the system's gravitational field in the same way as other types of energy. Hence, the introduction of gravitational self-energy to the Poisson equation is not ad hoc aiming at fitting the observational data. The results and further analysis with this modified Poisson equation agree with the observations without invoking dark matter. The gravitational self-energy density also influences the gravitational field along with the matter density $\rho$, the Poisson's equation thus takes the form,

$$\nabla^2\phi = 4\pi G\rho + K(\nabla\phi)^2 + \Lambda c^2 \quad (1)$$

where, $\phi \left(=\frac{GM}{r}\right)$ is the gravitational potential and the constant $K \approx \left(\frac{G^2}{c^2}\right)$. $K(\nabla\phi)^2$ is the gravitational self-energy density [23]. There are various density profiles for baryonic matter distribution in galaxies that influence the formation and evolution of a galaxy type. For instance, the De Vaucouleus profile for elliptical galaxies is characterized by a core radius and an exponential decline in density [24]. For disc galaxies, the exponential profile is common, where the stellar density decreases exponentially with distance from the centre. In this work, the assumed density distribution is such that it produces the observed rotational curves of galaxies. For regions within the core radius $r_0$, the density is a constant $\rho_0$, this gives a velocity that varies linearly with distance. Beyond $r_0$, the density falls off as [25],

$$\rho = \rho_o \left(\frac{r_o}{r}\right)^2 \qquad (2)$$

Equation (2) gives rise to a velocity that is independent of $r$, thereby accounting for the flat rotational curves of galaxies. This is also in accordance to the distribution of atomic hydrogen (HI) and molecular gas that often shows peak concentrations in the inner regions, tapering off outward [26]. The variation of $\rho$ with distance is shown in figure 1.

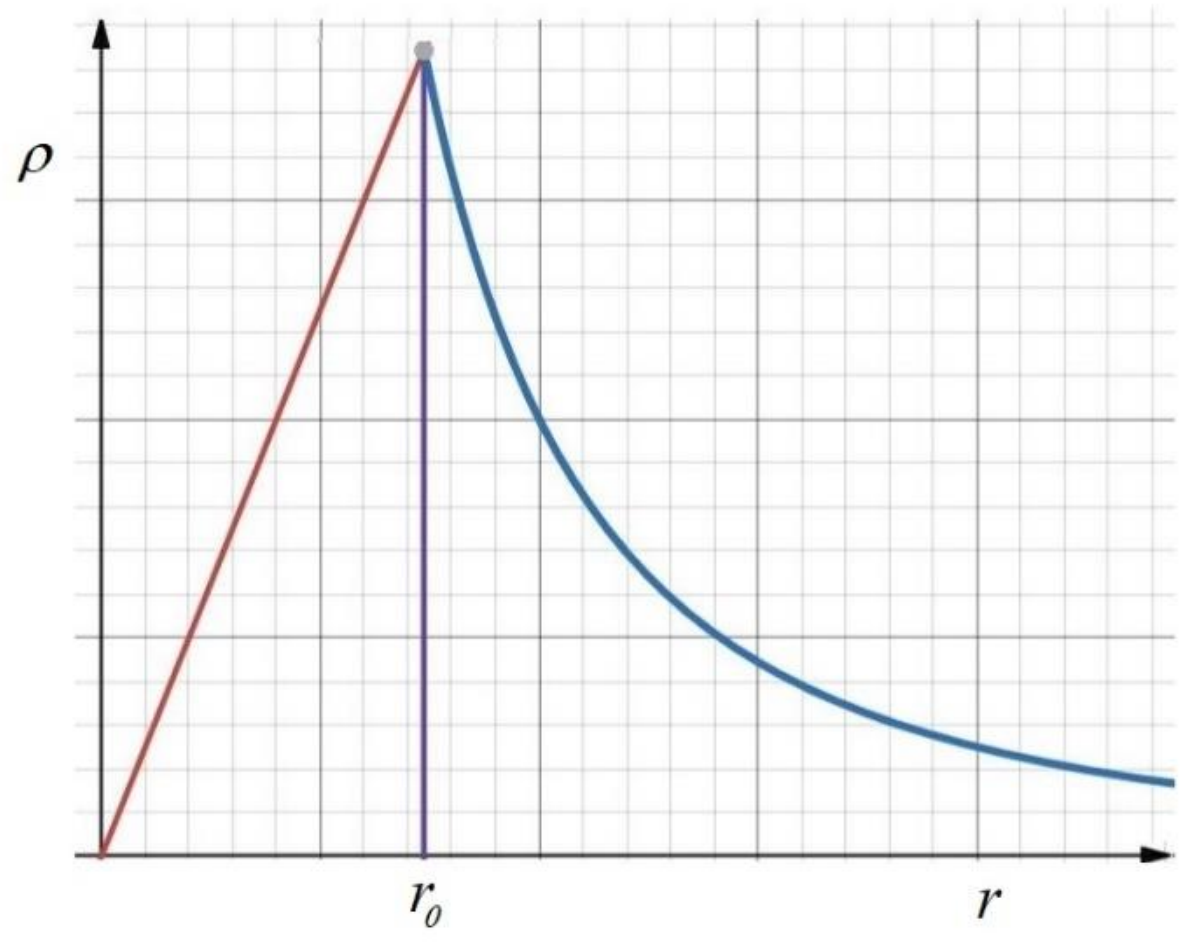

*Figure 1 Density versus Distance from the centre of a Galaxy*

For regions within the radius $r_0$, the matter density is uniform and dominant and the standard Poisson's equation takes its usual form (i.e., $\nabla^2\phi = 4\pi G\rho$) yielding the solution $\phi = \frac{GM}{r}$, which results in a velocity that varies linearly with distance. For regions farther away from the galactic centre, $> r_0$ , the matter density decreases according to Equation (2), and gravitational self-energy begins to dominate. At this juncture, equation (1) transforms into:

$$\nabla^2\phi = 4\pi G\rho_0 \left(\frac{r_0}{r}\right)^2 + K(\nabla\phi)^2 \qquad (3)$$

Here we neglect the cosmological constant term for the systems under consideration in this work as it does not play a role at the scale of individual galaxies or clusters.

The solution of equation (3) yields:

$$\phi = (Q + K') \ln \frac{r}{r_o} \tag{4}$$

where $Q = 4\pi G \rho_o {r_o}^2$ and $K' \approx \frac{GM}{r_o}$ are constants.

Equation (4) gives a force (per unit mass) of the form,

$$F = \frac{K"}{r} \tag{5}$$

where $K" = (GMa_{min})^{1/2}$ is also a constant.

Here $a_{min}$ is the minimum possible field strength (corresponding to the maximum size to which a gravitationally bound structure could grow).

$$a_{min} = \frac{GM}{r_{max}^2} \tag{6}$$

The outskirts of a large-scale structure would be governed by minimum gravitational field, characterized by the minimal acceleration $a_{min}$ as described in equation (6). The correction of gravity at low acceleration could also be a consequence of the quantum vacuum [27,28,29,30].

The value of the minimum acceleration is arrived from the requirement that large scale structures must have an attractive gravitational binding self-energy density $\frac{GM^2}{8\pi R^4}$ that at least equals the background repulsive dark energy density $\frac{\Lambda c^4}{8\pi G}$

$$\frac{GM^2}{8\pi R^4} = \frac{\Lambda c^4}{8\pi G} \tag{7}$$

where, $M$ is the total mass of the large-scale structure and $R$ is its observed size.

Equation (7) thus implies a mass-radius relation of the type,

$$\frac{M}{R^2} = \frac{c^2}{G}\sqrt{\Lambda} \approx 1\ g/cm^2 \tag{8}$$

From equation (6) and equation (8) we get,

$$\frac{a_{min}}{G} = \frac{c^2}{G}\sqrt{\Lambda} \tag{9}$$

Thereby $a_{min} \approx 10^{-8} cms^{-2}$, which is same as the Milgrom constant $a_0$.

The balance between the centripetal force and gravitational force gives [23],

$$v = (GMa_{min})^{1/4} \tag{10}$$

Equation (10) is similar to the mass asymptotic speed relation in MOND, with $a_{min}$ replacing the Milgrom constant $a_0$. Thus, in this model we have removed the ad-hoc nature $a_0$ by introducing a minimum possible field strength.

The dark energy term in equation (1) begins to dominate at regions much farther away from the galactic center (i.e. $\gg r_o$ ) as the matter density in these regions is very small. Equation (1) now takes the form,

$$\nabla^2\phi = K(\nabla\phi)^2 + \Lambda c^2 \quad (11)$$

where, $\Lambda\ (= 10^{-56} cm^{-2})$ is the constant cosmological constant term.

The solution of the above equation yields

$$\phi = A \ln secBr \quad (12)$$

where $A \approx (GMa_{min})^{1/2}$ and $B \approx \sqrt{\Lambda}c(GMa_{min})^{-1/4}$ are constants. This gives a force (per unit mass) of the form

$$F = \nabla\phi = B' tanBr \quad (13)$$

where $B' \approx \sqrt{\Lambda}c(GMa_{min})^{1/4}$ is also a constant.

On expanding equation (13) we get,

$$F = \nabla\phi = B'\left(Br + \frac{1}{3}(Br)^3 + \cdots\right) \quad (14)$$

The constants for a galaxy, such as the Andromeda with $M \approx 10^{45} g$, are $B \approx 10^{-25} cm^{-1}$ and $B' \approx 10^{-9} cms^{-11}$. Hence, the higher order terms in equation (14) become insignificant at greater distances (the galaxy's periphery). The constant $B'$ has little effect on the mass of the galaxy cluster; a two-order increase in mass would still result in the same value. This yields a force of the following form (per unit mass):

$$F = \Lambda c^2 r \quad (15)$$

The balance of the centripetal force and (modified) gravitational force now gives,

$$\frac{V^2}{r} = \Lambda c^2 r \quad (16)$$

The distance from the center of the structure at which the DE term starts to dominate can be estimated from Eq. (16). The balance between gravitational self energy and the repulsive DE gives,

$$\frac{(GMa_{min})^{\frac{1}{2}}}{r} = \Lambda c^2 r \quad (17)$$

This gives,

$$r = \sqrt{\frac{(GMa_{min})^{1/2}}{\Lambda c^2}} \quad (18)$$

Equation (18) can be used to constrain sizes to which galaxies can grow under the influence of gravity. Table 1 shows the observed size and the limit deduced from equation (18). The observed sizes of galaxies in the data set are well within the limit.

**Table 1:** Comparison of the observed size and the size limit set by MONG

(*Full data table is available in Appendix A*)

| *Galaxy* | *Mass* ($M_\odot$) | *R* (*kpc*) | *R (limit)* (*kpc*) |
|---|---|---|---|
| *M74* | 3E+11 | 13.9 | 2712 |
| *M99* | 1E+11 | 15.03 | 2061 |
| *NGC 2997* | 1E+11 | 18.34 | 2061 |
| *NGC 3370* | 1E+11 | 5.99 | 2050 |
| *NGC 5033* | 2E+12 | 27. 54 | 4375 |
| *NGC 4414* | 1E+11 | 12.69 | 1950 |
| *M33* | 5E+10 | 9.365 | 1733 |
| *NGC 5474* | 2.4E+10 | 4.21 | 1458 |
| *NGC 300* | 2.9E+10 | 14.42 | 1513 |
| *NGC 7793* | 1.39E+10 | 7.81 | 1260 |
| *M65* | 2E+12 | 17.37 | 2453 |
| *M96* | 7.99E+10 | 11.9 | 1950 |

## 3. Implications of Modification of Newtonian Gravity

### 3.1. Hot Gas in Clusters

Baryons, in the form of gas clouds in galaxy and galaxy cluster cool down sufficiently leading to the formation of stars. The baryons that do not form stars aggregate into hot gas in the intergalactic medium within galaxy clusters [31,32]. This region of hot gas is termed as inter cluster medium (ICM).

According to models of structure formation in the $\Lambda CDM$ model, the ICM must have cooled down at early epochs, enabling formation of new galaxies, hence the detection of extremely high temperatures of the ICM initially came as a surprise. The notion that X-ray emission from clusters were driven by diffuse intra-cluster gas at a temperature of $10^8 K$ and an atomic density of $n \approx 10^{-3} cm^{-3}$ was first proposed by [33]. The very first detected source of X-ray emissions from outside the Milky Way was M87, located at the centre of the Virgo Cluster [34,35]. Subsequently, within the next five years, other X-ray sources were identified in the Coma and Perseus clusters [36,37,38,39]. The Uhuru X-ray satellite which was launched to survey the X-ray emissions across the sky, found several clusters emitting strong X-rays, with luminosities ranging from $10^{36-38} W$ [40]. Cluster x-ray spectra are best explained by thermal bremsstrahlung from hot gas, suggesting that the space between galaxies within clusters is filled with very hot ($10^8 K$), low density gas ($n \approx 10 \quad -3cm \quad -3$) [41, 42].

The ICM makes up for around 10% of the cluster mass and consists mostly of ionised hydrogen and helium. The mass of ICM is found to be equivalent or sometimes even greater than the mass of all the galaxies within the cluster [43, 44, 45]. The estimated mass of the ICM in the Virgo cluster is $3 \times 10^{14} M_{\odot}$. This hot gas corresponds to a temperature of the order of $10^{8}K$ or even higher. At such a high temperature, the gas will undergo ionization and the free electrons will emit Bremsstrahlung radiations [45]. From equipartition of energy, the ions in the ICM will move with a velocity of,

$$v = \sqrt{\frac{KT}{m_p}} \approx 10^{8} cms^{-1} \tag{19}$$

where, $K$ is the Boltzmann's constant $(1.3807 \times 10^{-16} cm^{2} gs^{-1} K^{-1})$, $m_p$ is the mass of a proton and $T \sim 10^{8}K$

Conventionally, presence of hot gas within galaxy clusters serves as evidence supporting the presence of dark matter in the universe. The gas would experience the same gravitational force as the galaxies within the cluster, as a result, particle velocities comparable to the velocity dispersion in the cluster are required to prevent the gas from collapsing toward the centre of the cluster. But, the mass of all the visible matter in the galaxy cluster is insufficient to sustain the velocity of ions in the ICM and contain the hot gas. This discrepancy led to the inference that dark matter provides the additional gravitational force needed to balance the pressure of the hot gas. Consequently, galaxies must possess approximately five times more mass in the form of dark matter than what is visible, thus providing the requisite gravitational potential to retain the hot gas within the cluster.

Earlier preliminary work [46] indicated that the existence of hot gas in clusters can alternatively be explained without invoking dark matter through MONG. At the regions in the outskirts of galaxy clusters, where the hot gas is observed, the gravitational self-energy term indicated in equation (1) is more dominant. The extra term in MONG corresponding to gravitational self-energy will give a velocity of

$$v = (GMa_0)^{1/4} ln\left(\frac{r}{r_o}\right) \tag{20}$$

where, $M$ is the mass of the cluster, $a_0 \approx 10^{-8}\, cms^{2}$, $r_o$ is the distance at which the gravitational self-energy term begins to dominate. For a typical cluster like the Virgo cluster with mass of $M = 1.99M_{\odot}$, equation (20) gives a velocity of $v = 10^{8}cms^{-2}$. It is interesting to note that this velocity is same as the velocity of ions in the hot gas. Therefore, the predicted increase in the gravitational force in MONG (as given by equation (5) provides the required velocity to hold ions of the hot gas in place. Therefore, the logarithmical rise in potential with distance provides the necessary gravitational potential to retain the gas within the cluster (for example, for $r$ of the order of 10 $r_c$ this implies an order of about 3 times increase in force).

This eliminates the necessity for dark matter within the clusters. Table 2 compares the Newtonian velocity of ions in the ICM with the MONG velocities. We utilized data from [47] and from [48]. These datasets provide crucial observational parameters, including mass, radius, and temperature, which were essential for this analysis. The value of $M$ is taken as the gas mass $M_{gas}$, and $R_{500}$ is the radius enclosing 500 times the critical density of the sources. The parameter $r_o$ is calculated using the formula: $r_o = \sqrt{GM}/a_0$ where $M$ is $M_{gas}$, within $R_{500}$ deduced from the catalogue. The observational data were meticulously integrated into our model to conduct a comprehensive analysis and thereby facilitate a thorough comparison of the theory. It is seen that the two velocities are indeed of the same order, thus implying that the gravitational force from MONG provides the required potential well for such high temperature gas without invoking dark matter.

**Table 2:** Calculated velocities using Newtonian velocity & MONG velocity

(*Full data table is available in the Appendix B*)

| *Cluster* | *z* | *kT (keV)* | *R500 (kpc)* | *$M_{gas}$ ($M_\odot$)* | *Ion velocity (m/s) +/- Error (m/s)* | *v(MONG) (m/s)* |
|---|---|---|---|---|---|---|
| Abell119 | 0.044 | 5.79 | 1.13E+03 | 4.50E+13 | 7.45E+05 +/- 2.63E+05 | 5.75E+05 |
| Abell1413 | 0.1427 | 8.42 | 1.31E+03 | 8.11E+13 | 8.98E+05 +/- 3.17E+05 | 6.02E+05 |
| HydraA | 0.0538 | 3.75 | 1.02E+03 | 4.20E+13 | 5.99E+05 +/- 2.11E+05 | 5.40E+05 |
| J001639.2-010209 | 0.1744 | 1.55 | 5.04E+02 | 4.32E+12 | 3.85E+05 +/- 1.36E+05 | 3.98E+05 |
| J001737.3-005239 | 0.2065 | 4.34 | 8.57E+02 | 2.20E+13 | 6.45E+05 +/- 2.28E+05 | 5.07E+05 |
| MKW3S | 0.045 | 3.44 | 9.09E+02 | 2.35E+13 | 5.74E+05 +/- 2.03E+05 | 5.24E+05 |
| MKW8 | 0.027 | 2.5 | 7.16E+02 | 1.13E+13 | 4.89E+05 +/- 1.73E+05 | 4.71E+05 |
| PKS0745-191 | 0.1028 | 6.76 | 1.35E+03 | 8.09E+13 | 8.05E+05 +/- 2.84E+05 | 6.15E+05 |
| UGC3957 | 0.034 | 2.34 | 6.97E+02 | 1.05E+13 | 4.73E+05 +/- 1.67E+05 | 4.65E+05 |
| ZwCl1215+0400 | 0.075 | 7.57 | 1.27E+03 | 6.57E+13 | 8.52E+05 +/- 3.01E+05 | 6.02E+05 |

## 3.2 Angular momentum of Galaxies

Angular momentum per unit mass, or stellar specific angular momentum, is a crucial factor in the formation and evolution of galaxies [49,50,51,52,53]. It is discovered that the stellar specific angular momentum and the stellar mass of both, nearby and distant galaxies are related by the fundamental relation [54,55,56,57]

$$j \propto M^{2/3} \tag{21}$$

In the standard ΛCDM model, equation (21) describes the coupling between the baryonic matter in galaxies and their dark matter haloes. In particular, the specific angular momentum is dependent on the 2/3 power of the mass due to tidal torques [58, 59]. A recent work [60] studied the scaling relation for a large sample of galaxies and concluded that the power law remains unbroken with a slope of 0.5–0.6. Various galaxy models are successful in arriving at the relation for massive spiral and disc galaxies [61, 62, 63]. However, there a discrepancy in the relation for lower mass galaxies [64, 65, 66] for which few models predict flattening of the relation whereas others do not show no such change in comparison with large spirals. The disparity is thought to be caused by a dearth of observational estimates over a wide range of galaxy star masses.

The ratios of stellar to halo $j_*/j_h$ and baryonic to halo $j_b/j_h$ measures the level of conservation of specific angular momentum [67]. Here, $j_h$ is not directly observable, but derived through the relation $j_h = {M_h}^{2/3}$ [53]. The ratios fall below unity, aligning with the predictions of conventional disk formation models [68, 52, 56, 53]. The specific angular momentum is also important in explaining galaxy morphology. While galaxies with lower angular momentum are spheroidal or elliptical, galaxies with higher angular momentum tend to display extended and flattened characteristics such as spiral arms and disks. Early-type galaxies are found to have lower specific angular momentum that late-type galaxies [68, 53]. But recent results show that early type galaxies are not just ellipticals. Though the specific angular momentum plays a role in galaxy morphology, it is not the only deciding factor. There is a significant correlation between galaxy dynamics, their stellar mass and the environment [69, 70].

In alternate models of gravity, dark matter halos will have no contribution to the angular momentum of the galaxy. Hence the angular momentum is purely from the baryonic component. Substituting for $R$ from equation (1) and $v$ from equation (8) in the usual expression for angular momentum $(J = MVR)$ we get,

$$J = \frac{1}{c}\left(\frac{G^3 a_{min}}{\Lambda}\right)^{1/4} M^{1.75} \tag{22}$$

Where the constants have their usual values, which gives:

$$J = 1.38 \times 10^{-4} M^{1.75}\ ergs \tag{23}$$

The angular momentum dependence on mass is seen to have increased in the case of MONG in comparison to ΛCDM model. This is expected as there is no dark matter component. The angular momentum estimated from MONG for galaxies such as the Milky Way $(J \approx 10^{73}\ ergs)$ and Andromeda $(J \approx 10^{74}\ ergs)$ are same as that observed.

The Spritzer Photometry and Accurate Rotation Curves (SPARC) database was used to examine the angular momentum arrived from MONG $(J_{MONG})$. With its new surface photometry at $3.6\,\mu m$ and rotation curves generated from previous $H\,I/H\alpha$ studies, the SPARC provides data for about 175 nearby galaxies. It stands out as a particularly comprehensive dataset due to the large range of galaxy morphologies and luminosities $(\sim 5\ dex)$. Using a combination of high quality $H\,I$ rotation curves, $H\,I$ surface densities, and near-infrared surface brightness profiles, a dataset was constructed to understand the contribution of stars and gas to the total angular momentum of galaxies [71]. [56] contains data on significant galaxy attributes including total baryonic mass, gas mass, stellar mass and specific angular momenta. The halo mass data was taken from [72], and LITTLE THINGS galaxies from [73] through techniques involving rotation curve decomposition.

Fig (1) shows the scaling relation between the baryon angular momentum $J_B$ and $J_{MONG}$, with a slope $(J_B\ /\ J_{MONG})$ of 0.77 which is in very good accordance with the scaling relation between $J_B$ and $J_h$. Scaling relation between the stellar angular momentum $J_*$ and $J_{MONG}$ plotted in fig (2) has a slope $(J_*/J_{MONG})$ of 0.85 with a corelation of 0.95, while the stellar to halo ratio $J_*/J_h\ = 0.75$ with a corelation of 0.79.

Equations (4 – 6) from [72] are utilized to compute the halo mass in order to make a comparison between $J_{MONG}$ and $J_H$. The correlation ratio is determined through a Chi-square fit technique and plotted in figure (3). The correlation coefficient is almost unity, showing a replicable result.

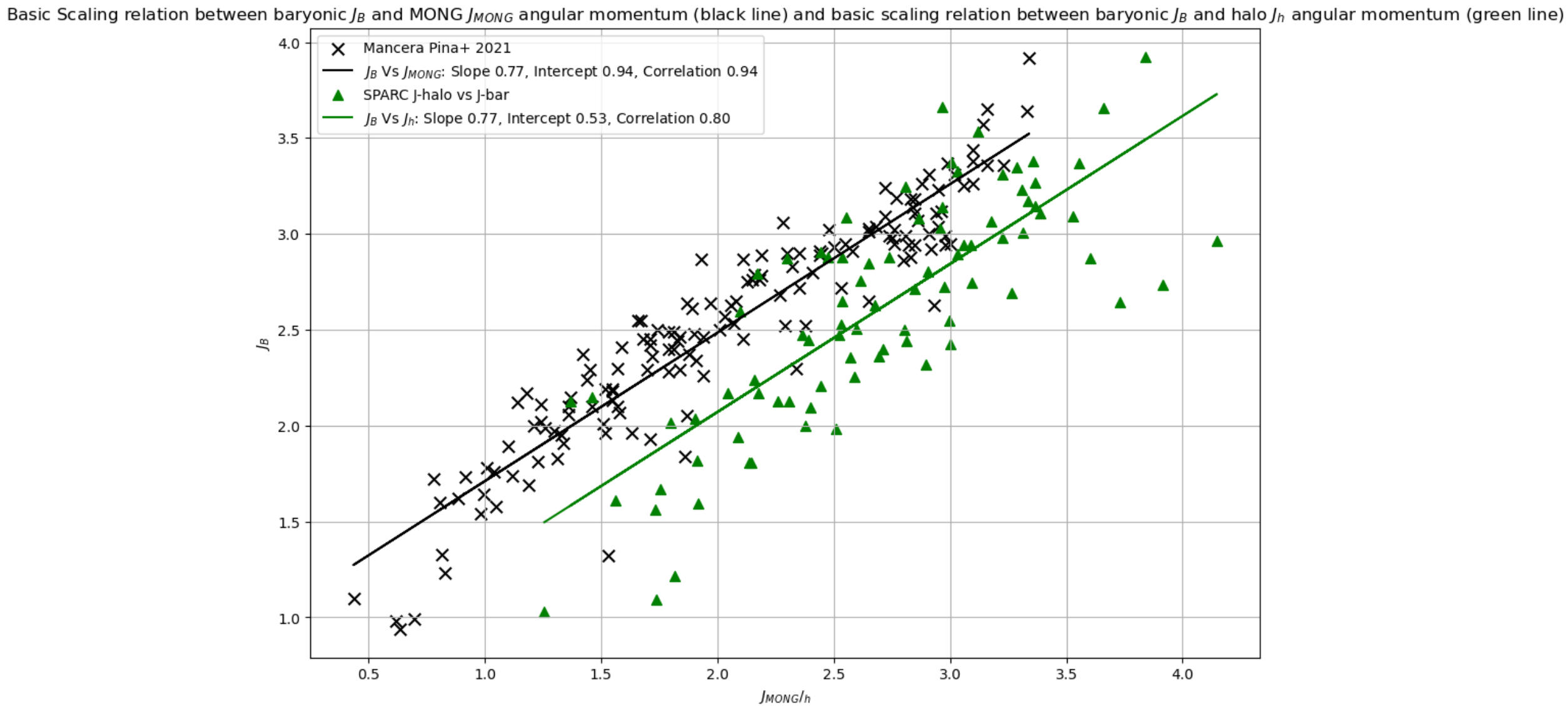

***Figure 2:*** *Basic scaling relation between baryonic $J_B$ and MONG $J_{MONG}$ angular momentum (black line) and basic scaling relation between baryonic $J_B$ and halo $J_h$ angular momentum (green line)*

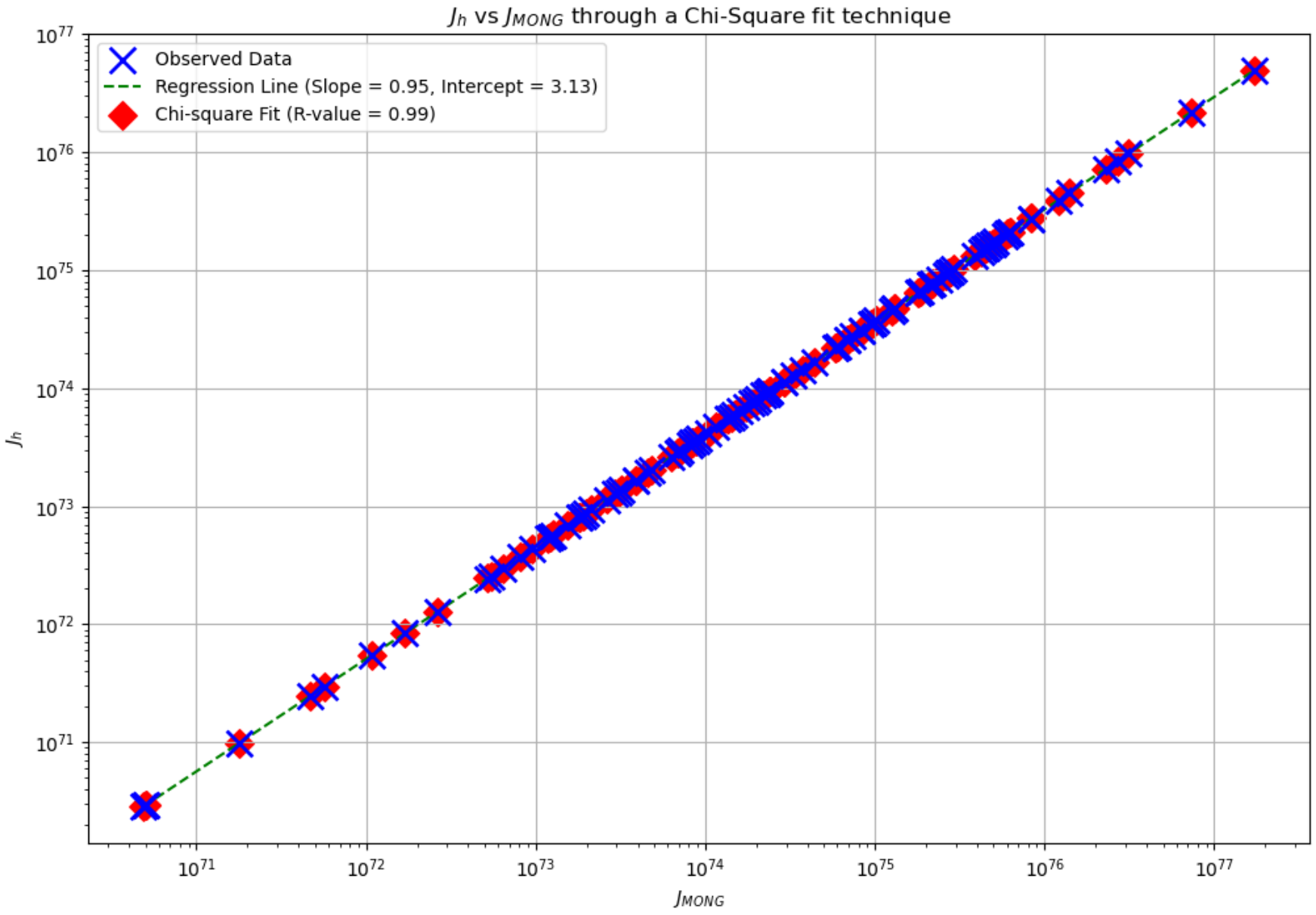

***Figure 3:*** *$J_h$ versus $J_{MONG}$ through a Chi-square fit technique*

## 4. Conclusion

Direct detection of dark matter has been elusive for decades in spite of bigger and advances detectors. And as a result of these negative results alternate models of gravity have been introduced to explain observations typically attributed to dark matter. First of these was the Modification of Newtonian Dynamics (MOND), which introduced an ad-hoc minimal acceleration to explain flat rotation curves of spirals galaxies. The original MOND theory has evolved to incorporate relativistic aspects to explain cosmological phenomena like gravitational lensing.

An alternate model, that does not invoke an ad-hoc acceleration, is by modifying Newtonian gravity. This model (MONG) was demonstrated to be successful in explaining the flat rotation curves of spiral galaxies, velocity dispersion of galactic clusters, infall of galaxies in Virgo cluster, etc. Here we have extended MONG to explain the high velocity of hot gas in clusters without the need to invoke dark matter. As shown, in the outskirts of galaxy clusters, where the hot gas is observed, the gravitational self-energy term becomes more dominant. As a result of this increased gravitational strength, the cluster is able to hold this hot gas, without the need for dark matter.

We also see how MONG gives a galactic angular momentum with a higher dependence on mass ($M^{1.75}$) as opposed to the $M^{2/3}$ in Newtonian. This higher dependence accounts for the observed angular momenta of galaxies without the need for dark matter. These results from MONG are consistent with observations.

In a recent study [74], it was demonstrated that a violation of the Equivalence Principle at large distances can lead to the rediscovery of MOND-like theories when the approach to gravitation adheres to the Mach Principle. MOND-like theories are retrieved at low energies by fixing a nongeodesic ratio between the inertial and gravitational masses by considering a direct coupling between the Ricci curvature scalar and the matter Lagrangian. Studies have also investigated if, in accordance with Mach's principle, the gravitational interaction energy of the background quantum vacuum energy function as a global Higg's field and contribute to the local inertial masses of particles [75].

# Appendix

## A. Comparison of the observed size of galaxies and the size limit set by MONG

| Galaxy | Mass ($M_\odot$) | R (kpc) | R (limit) (kpc) |
|---|---|---|---|
| **M104,** | 7.9E+11 | 161.5780224 | 34664.61703 |
| **M64** | 4.02E+8 | 82.5387732 | 5183.569282 |
| **M94** | 5.9E+10 | 104.485788 | 18140.60314 |
| **M81** | 6.67E+11 | 147.1799808 | 33161.38116 |
| **M77** | 9.99E+11 | 138.481164 | 36653.3699 |
| **NGC 2841** | 6.9E+10 | 227.9189988 | 18853.34521 |
| **NGC 7742** | 3.21E+7 | 77.739426 | 3466.461703 |
| **M88** | 9.8E+10 | 201.2726232 | 25917.84641 |
| **M31,** | 1.49E+12 | 232.8683256 | 40563.62177 |
| **NGC 891** | 6.36E+10 | 60.991704 | 27567.75525 |
| **NGC 3184** | 1.87E+13 | 164.1776688 | 73306.7398 |
| **NGC 7331** | 1.62E+12 | 224.219502 | 39869.9709 |
| **M51a** | 1.6E+11 | 158.1784848 | 23181.62656 |
| **M63** | 1.39E+11 | 168.227118 | 22420.53227 |
| **M74** | 2.99E+11 | 130.7822112 | 27126.52879 |
| **M99** | 9.99E+10 | 150.479532 | 20611.7046 |
| **NGC 2997** | 9.99E+10 | 183.225078 | 20611.7046 |
| **NGC 3370** | 9.99E+10 | 60.0918264 | 20611.7046 |
| **NGC 5033** | 1.98E+12 | 275.4125388 | 43588.44828 |
| **NGC 4414** | 7.9E+10 | 126.4328028 | 19493.34668 |
| **M33** | 2.9E+11 | 93.73725 | 17332.30851 |
| **NGC 5474** | 2.49E+10 | 42.1442676 | 14574.6761 |
| **NGC 300** | 2.89E+10 | 144.188748 | 15125.62779 |
| **NGC 7793** | 13.9E+10 | 78.0393852 | 12607.99183 |
| **M65** | 1.98E+11 | 173.7763632 | 24511.58577 |
| **M96** | 3.30E+10 | 119.6337276 | 19493.34668 |
| **M98** | 1.98E+11 | 214.3208484 | 24511.58577 |
| **M58** | 2.99E+11 | 203.5723104 | 27126.52879 |
| **M66** | 1.30E+11 | 130.5822384 | 22008.97149 |
| **NGC 4216** | 1.99E+10 | 249.466068 | 6517.993301 |
| **NGC 3310** | 2.19E+10 | 72.3901536 | 14116.25776 |
| **M61** | 6.95E+10 | 141.480756 | 18853.34521 |
| **Maffei 2** | 2.48E+9 | 82.8387324 | 8195.942671 |
| **NGC 5005** | 6.9E+10 | 192.973752 | 18819.58797 |
| **NGC 7752** | 1.79E+10 | 101.4362028 | 13425.54844 |
| **M100** | 1.99E+11 | 254.6153676 | 24511.58577 |
| **NGC 3521** | 4.97E+10 | 223.71957 | 17332.30851 |
| **NGC 1087** | 2. 68 E+10 | 81.3889296 | 14857.81173 |
| **NGC 6872** | 1.30E+10 | 1098.000652 | 6959.847886 |
| **NGC 253** | 1.00077E+11 | 184.7748672 | 20611.7046 |
| **NGC 6946** | 7.29204E+11 | 133.8317964 | 33880.09542 |
| **NGC 2403** | 1.00077E+11 | 138.4311708 | 20611.7046 |
| **NGC 4625** | 9.80 E+9 | 47.993472 | 11532.41942 |
| **NGC 1512** | 3.00231E+11 | 326.8555416 | 27126.52879 |
| **NGC 1365** | 3.60076E+11 | 337.1541408 | 28391.57824 |
| **M95** | 5.00 E+10 | 122.8832856 | 17332.30851 |
| **NGC 1300** | 4.90 E+10 | 196.973208 | 17244.98932 |
| **M109** | 2.49941E+11 | 270.1632528 | 25917.84641 |
| **Milky Way** | 1.15164E+12 | 133.981776 | 37956.69323 |
| **NGC 1090** | 1.10135E+11 | 178.0757784 | 21108.72884 |
| **NGC 3079** | 1.20 E+6 | 207.8717256 | 1213.135247 |
| **NGC 7479** | 2.29825E+11 | 180.975384 | 25383.1706 |
| **UGC 10214** | 1.28239E+11 | 855.1836792 | 21923.8288 |
| **M108** | 1.25222E+11 | 28.9460628 | 15379.92909 |
| **NGC 4945** | 1.39806E+11 | 145.5302052 | 21794.22414 |
| **NGC 4631** | 1.36 E+10 | 253.3155444 | 22420.53227 |
| **NGC 4618** | 1 E+9 | 125.982864 | 12516.95344 |
| **NGC 3109** | 2.29 E+9 | 53.6926968 | 6517.993301 |

## B. Comparison between the Ion velocities and MONG velocity

| ID | z | kT (keV) | R500 (cm) | $M_{gas}$ ($M_\odot$) | Ion velocity (m/s) | v_MONG (m/s) | Reference |
|---|---|---|---|---|---|---|---|
| **Abell119** | 0.044 | 5.79 | 1.33E+04 | 4.50E+13 | 7.45E+05 +/- 2.63E+05 | 5.75E+05 | Elkholy+, 2015 |
| **Abell1413** | 0.1427 | 8.42 | 1.57E+04 | 8.11E+13 | 8.98E+05 +/- 3.17E+05 | 6.02E+05 | Elkholy+, 2015 |
| **Abell1644** | 0.0474 | 5.31 | 1.45E+04 | 5.50E+13 | 7.13E+05 +/- 2.52E+05 | 5.72E+05 | Elkholy+, 2015 |
| **Abell1651** | 0.086 | 7.52 | 1.45E+04 | 7.30E+13 | 8.49E+05 +/- 3.00E+05 | 5.99E+05 | Elkholy+, 2015 |
| **Abell1689** | 0.184 | 10.4 | 1.58E+04 | 1.24E+14 | 9.98E+05 +/- 3.52E+05 | 6.15E+05 | Elkholy+, 2015 |
| **Abell1736** | 0.0461 | 3.19 | 1.47E+04 | 2.80E+13 | 5.53E+05 +/- 1.95E+05 | 5.22E+05 | Elkholy+, 2015 |
| **Abell1795** | 0.0616 | 6.75 | 1.30E+04 | 6.19E+13 | 8.04E+05 +/- 2.84E+05 | 5.92E+05 | Elkholy+, 2015 |
| **Abell1914** | 0.1712 | 10.1 | 1.31E+04 | 1.23E+14 | 9.84E+05 +/- 3.47E+05 | 6.15E+05 | Elkholy+, 2015 |
| **Abell2029** | 0.0767 | 8.2 | 1.44E+04 | 1.25E+14 | 8.86E+05 +/- 3.13E+05 | 6.10E+05 | Elkholy+, 2015 |
| **Abell2063** | 0.0354 | 3.45 | 1.33E+04 | 2.36E+13 | 5.75E+05 +/- 2.03E+05 | 5.23E+05 | Elkholy+, 2015 |
| **Abell2065** | 0.0721 | 6.18 | 1.37E+04 | 5.80E+13 | 7.69E+05 +/- 2.71E+05 | 5.82E+05 | Elkholy+, 2015 |
| **Abell2142** | 0.0899 | 11.2 | 1.35E+04 | 1.45E+14 | 1.04E+06 +/- 3.67E+05 | 6.40E+05 | Elkholy+, 2015 |
| **Abell2147** | 0.0351 | 4.56 | 1.39E+04 | 3.44E+13 | 6.61E+05 +/- 2.33E+05 | 5.53E+05 | Elkholy+, 2015 |
| **Abell2204** | 0.1523 | 9.47 | 1.27E+04 | 1.36E+14 | 9.52E+05 +/- 3.36E+05 | 6.13E+05 | Elkholy+, 2015 |
| **Abell2244** | 0.097 | 5.88 | 1.28E+04 | 6.04E+13 | 7.50E+05 +/- 2.65E+05 | 5.77E+05 | Elkholy+, 2015 |
| **Abell2256** | 0.0601 | 7.73 | 1.35E+04 | 9.54E+13 | 8.60E+05 +/- 3.04E+05 | 6.10E+05 | Elkholy+, 2015 |
| **Abell2319** | 0.0564 | 10.1 | 1.23E+04 | 1.56E+14 | 9.84E+05 +/- 3.47E+05 | 6.36E+05 | Elkholy+, 2015 |
| **Abell2657** | 0.0404 | 3.82 | 1.32E+04 | 2.08E+13 | 6.05E+05 +/- 2.14E+05 | 5.25E+05 | Elkholy+, 2015 |
| **Abell2734** | 0.062 | 4.28 | 1.26E+04 | 3.02E+13 | 6.40E+05 +/- 2.26E+05 | 5.41E+05 | Elkholy+, 2015 |
| **Abell3112** | 0.075 | 5.01 | 1.19E+04 | 4.81E+13 | 6.93E+05 +/- 2.45E+05 | 5.58E+05 | Elkholy+, 2015 |
| **Abell3158** | 0.059 | 5.2 | 1.18E+04 | 4.53E+13 | 7.06E+05 +/- 2.49E+05 | 5.65E+05 | Elkholy+, 2015 |
| **Abell3376** | 0.0455 | 4.58 | 1.18E+04 | 2.49E+13 | 6.62E+05 +/- 2.34E+05 | 5.41E+05 | Elkholy+, 2015 |
| **Abell3391** | 0.0531 | 5.41 | 1.13E+04 | 4.28E+13 | 7.20E+05 +/- 2.54E+05 | 5.67E+05 | Elkholy+, 2015 |
| **Abell3571** | 0.0397 | 7.73 | 1.10E+04 | 8.43E+13 | 8.60E+05 +/- 3.04E+05 | 6.12E+05 | Elkholy+, 2015 |
| **Abell3667** | 0.056 | 6.6 | 1.15E+04 | 9.38E+13 | 7.95E+05 +/- 2.81E+05 | 5.94E+05 | Elkholy+, 2015 |
| **Abell3822** | 0.076 | 5.36 | 1.15E+04 | 5.57E+13 | 7.17E+05 +/- 2.53E+05 | 5.71E+05 | Elkholy+, 2015 |
| **Abell3827** | 0.098 | 7.89 | 1.10E+04 | 8.46E+13 | 8.69E+05 +/- 3.07E+05 | 6.05E+05 | Elkholy+, 2015 |
| **Abell3921** | 0.0936 | 5.93 | 1.09E+04 | 6.15E+13 | 7.54E+05 +/- 2.66E+05 | 5.75E+05 | Elkholy+, 2015 |
| **Abell399** | 0.0715 | 6.47 | 9.86E+03 | 7.43E+13 | 7.87E+05 +/- 2.78E+05 | 5.91E+05 | Elkholy+, 2015 |
| **Abell400** | 0.024 | 2.15 | 9.65E+03 | 8.00E+12 | 4.54E+05 +/- 1.60E+05 | 4.61E+05 | Elkholy+, 2015 |
| **Abell4038** | 0.0283 | 3.12 | 1.03E+04 | 1.71E+13 | 5.47E+05 +/- 1.93E+05 | 5.07E+05 | Elkholy+, 2015 |
| **Abell4059** | 0.046 | 4.34 | 9.58E+03 | 2.41E+13 | 6.45E+05 +/- 2.28E+05 | 5.37E+05 | Elkholy+, 2015 |
| **Abell478** | 0.09 | 7.65 | 9.49E+03 | 1.13E+14 | 8.56E+05 +/- 3.02E+05 | 6.02E+05 | Elkholy+, 2015 |
| **Abell539** | 0.0288 | 2.59 | 9.83E+03 | 1.63E+13 | 4.98E+05 +/- 1.76E+05 | 4.96E+05 | Elkholy+, 2015 |
| **Abell644** | 0.0704 | 8.49 | 9.03E+03 | 6.86E+13 | 9.02E+05 +/- 3.18E+05 | 6.09E+05 | Elkholy+, 2015 |
| **Abell754** | 0.0528 | 11.8 | 1.02E+04 | 5.38E+13 | 1.06E+06 +/- 3.74E+05 | 6.31E+05 | Elkholy+, 2015 |
| **AbellS405** | 0.0613 | 4.62 | 9.07E+03 | 2.85E+13 | 6.65E+05 +/- 2.35E+05 | 5.44E+05 | Elkholy+, 2015 |
| **HydraA** | 0.0538 | 3.75 | 9.10E+03 | 4.20E+13 | 5.99E+05 +/- 2.11E+05 | 5.40E+05 | Elkholy+, 2015 |
| **J001639.2-010209** | 0.1744 | 1.55 | 9.21E+03 | 4.32E+12 | 3.85E+05 +/- 1.36E+05 | 3.98E+05 | Elkholy+, 2015 |
| **J001737.3-005239** | 0.2065 | 4.34 | 8.40E+03 | 2.20E+13 | 6.45E+05 +/- 2.28E+05 | 5.07E+05 | Elkholy+, 2015 |
| **J002223.6+001202** | 0.2791 | 1.14 | 8.04E+03 | 3.73E+12 | 3.30E+05 +/- 1.16E+05 | 3.81E+05 | Elkholy+, 2015 |
| **J002314.3+001158** | 0.2734 | 1.31 | 7.17E+03 | 4.45E+12 | 3.54E+05 +/- 1.25E+05 | 3.92E+05 | Elkholy+, 2015 |
| **J003917.9+004200** | 0.2801 | 1.07 | 6.98E+03 | 4.05E+12 | 3.20E+05 +/- 1.13E+05 | 3.85E+05 | Elkholy+, 2015 |
| **J003942.2+004533** | 0.4152 | 1.83 | 7.19E+03 | 8.04E+12 | 4.19E+05 +/- 1.48E+05 | 4.16E+05 | Elkholy+, 2015 |
| **J004231.2+005114** | 0.1468 | 1.41 | 6.78E+03 | 3.29E+12 | 3.68E+05 +/- 1.30E+05 | 3.82E+05 | Elkholy+, 2015 |
| **J004252.6+004259** | 0.2595 | 1.98 | 4.75E+03 | 5.68E+12 | 4.36E+05 +/- 1.54E+05 | 4.08E+05 | Takey+, 2011 |
| **J004253.7-093423** | 0.4054 | 1.49 | 5.67E+03 | 7.40E+12 | 3.78E+05 +/- 1.33E+05 | 4.12E+05 | Takey+, 2011 |
| **J004334.0+010106** | 0.1741 | 1.34 | 4.65E+03 | 5.61E+12 | 3.58E+05 +/- 1.26E+05 | 4.15E+05 | Takey+, 2011 |
| **J004350.7+004733** | 0.4754 | 2.32 | 5.36E+03 | 9.06E+12 | 4.71E+05 +/- 1.66E+05 | 4.18E+05 | Takey+, 2011 |
| **J004401.3+000647** | 0.2185 | 1.83 | 5.54E+03 | 8.52E+12 | 4.19E+05 +/- 1.48E+05 | 4.39E+05 | Takey+, 2011 |
| **J005259.6-083936** | 0.3155 | 1.7 | 5.50E+03 | 1.01E+13 | 4.04E+05 +/- 1.43E+05 | 4.41E+05 | Takey+, 2011 |
| **J015917.2+003011** | 0.2882 | 1.98 | 5.77E+03 | 2.26E+13 | 4.36E+05 +/- 1.54E+05 | 4.99E+05 | Takey+, 2011 |
| **J020342.0-074652** | 0.4398 | 3.71 | 6.23E+03 | 1.94E+13 | 5.96E+05 +/- 2.10E+05 | 4.71E+05 | Takey+, 2011 |
| **J021012.7-001451** | 0.2834 | 0.71 | 8.40E+03 | 3.06E+12 | 2.61E+05 +/- 9.21E+04 | 3.68E+05 | Takey+, 2011 |
| **J023150.5-072836** | 0.1791 | 0.65 | 7.53E+03 | 2.28E+12 | 2.50E+05 +/- 8.82E+04 | 3.58E+05 | Takey+, 2011 |
| **J023346.9-085054** | 0.2799 | 1.78 | 4.07E+03 | 6.70E+12 | 4.13E+05 +/- 1.46E+05 | 4.17E+05 | Takey+, 2011 |
| **J030637.1-001803** | 0.4576 | 2.05 | 5.62E+03 | 6.98E+12 | 4.43E+05 +/- 1.56E+05 | 4.04E+05 | Takey+, 2011 |
| **J033757.5+002900** | 0.3232 | 1.8 | 5.32E+03 | 7.22E+12 | 4.15E+05 +/- 1.46E+05 | 4.18E+05 | Takey+, 2011 |
| **J073605.9+433906** | 0.4282 | 3.87 | 5.67E+03 | 1.91E+13 | 6.09E+05 +/- 2.15E+05 | 4.71E+05 | Takey+, 2011 |
| **J075121.7+181600** | 0.3882 | 1.2 | 7.53E+03 | 5.39E+12 | 3.39E+05 +/- 1.20E+05 | 3.94E+05 | Takey+, 2011 |
| **J075427.4+220949** | 0.3969 | 2.18 | 5.02E+03 | 1.15E+13 | 4.57E+05 +/- 1.61E+05 | 4.41E+05 | Takey+, 2011 |
| **J075839.2+351936** | 0.1658 | 1.44 | 6.44E+03 | 4.31E+12 | 3.71E+05 +/- 1.31E+05 | 3.98E+05 | Takey+, 2011 |
| **J081058.2+500529** | 0.3972 | 2.05 | 5.31E+03 | 9.15E+12 | 4.43E+05 +/- 1.56E+05 | 4.26E+05 | Takey+, 2011 |
| **J082412.9+300437** | 0.3008 | 1.8 | 5.66E+03 | 6.49E+12 | 4.15E+05 +/- 1.46E+05 | 4.13E+05 | Takey+, 2011 |
| **J082746.9+263508** | 0.3869 | 1.69 | 1.10E+04 | 6.38E+12 | 4.02E+05 +/- 1.42E+05 | 4.05E+05 | Takey+, 2011 |
| **J083146.1+525056** | 0.5383 | 3.47 | 6.78E+03 | 9.23E+12 | 5.77E+05 +/- 2.04E+05 | 4.13E+05 | Takey+, 2011 |
| **J083454.8+553422** | 0.2421 | 3.4 | 5.83E+03 | 4.87E+13 | 5.71E+05 +/- 2.02E+05 | 5.62E+05 | Takey+, 2011 |
| **J083724.7+553249** | 0.2767 | 1.83 | 5.68E+03 | 1.17E+13 | 4.19E+05 +/- 1.48E+05 | 4.55E+05 | Takey+, 2011 |
| **J083727.0+145102** | 0.2154 | 2 | 7.14E+03 | 1.23E+13 | 4.38E+05 +/- 1.55E+05 | 4.64E+05 | Takey+, 2011 |
| **J084004.3+294544** | 0.5644 | 1.74 | 5.68E+03 | 7.55E+12 | 4.08E+05 +/- 1.44E+05 | 3.98E+05 | Takey+, 2011 |
| **J084105.6+383158** | 0.2313 | 1.1 | 4.24E+03 | 5.16E+12 | 3.25E+05 +/- 1.15E+05 | 4.05E+05 | Takey+, 2011 |
| **J084701.9+345114** | 0.4643 | 2.73 | 7.14E+03 | 9.27E+12 | 5.11E+05 +/- 1.80E+05 | 4.21E+05 | Takey+, 2011 |
| **J084847.8+445611** | 0.5753 | 1.92 | 4.77E+03 | 9.74E+12 | 4.29E+05 +/- 1.51E+05 | 4.12E+05 | Takey+, 2011 |
| **J085253.3+175718** | 0.2106 | 1.23 | 5.07E+03 | 3.37E+12 | 3.43E+05 +/- 1.21E+05 | 3.79E+05 | Takey+, 2011 |
| **J085332.1+150133** | 0.4298 | 2.63 | 7.57E+03 | 1.15E+13 | 5.02E+05 +/- 1.77E+05 | 4.38E+05 | Takey+, 2011 |
| **J085500.0+150918** | 0.2246 | 1.47 | 8.16E+03 | 3.34E+12 | 3.75E+05 +/- 1.32E+05 | 3.78E+05 | Takey+, 2011 |
| **J085757.3+170415** | 0.3228 | 1.34 | 7.27E+03 | 7.59E+12 | 3.58E+05 +/- 1.26E+05 | 4.22E+05 | Takey+, 2011 |
| **J091255.3+415553** | 0.152 | 0.68 | 4.30E+03 | 2.81E+12 | 2.55E+05 +/- 9.00E+04 | 3.72E+05 | Takey+, 2011 |
| **J091935.0+303157** | 0.4273 | 3.88 | 5.25E+03 | 2.23E+13 | 6.10E+05 +/- 2.15E+05 | 4.81E+05 | Takey+, 2011 |
| **J092539.8+362702** | 0.1152 | 2.67 | 5.17E+03 | 2.04E+13 | 5.06E+05 +/- 1.79E+05 | 5.11E+05 | Takey+, 2011 |
| **J092545.5+305858** | 0.5865 | 3.57 | 4.50E+03 | 1.96E+13 | 5.85E+05 +/- 2.06E+05 | 4.53E+05 | Takey+, 2011 |
| **J093205.0+473320** | 0.2248 | 1.36 | 4.59E+03 | 6.51E+12 | 3.61E+05 +/- 1.27E+05 | 4.20E+05 | Takey+, 2011 |
| **J093710.9+611558** | 0.2079 | 0.68 | 4.18E+03 | 2.66E+12 | 2.55E+05 +/- 9.00E+04 | 3.65E+05 | Takey+, 2011 |
| **J094054.6+032226** | 0.4928 | 3.26 | 6.59E+03 | 1.76E+13 | 5.59E+05 +/- 1.97E+05 | 4.58E+05 | Takey+, 2011 |
| **J094437.5+040035** | 0.3646 | 1.82 | 4.43E+03 | 4.51E+12 | 4.18E+05 +/- 1.48E+05 | 3.85E+05 | Takey+, 2011 |
| **J094530.9+094637** | 0.2114 | 1.44 | 6.53E+03 | 4.58E+12 | 3.71E+05 +/- 1.31E+05 | 3.99E+05 | Takey+, 2011 |
| **J094909.6+582307** | 0.2555 | 1.74 | 5.11E+03 | 6.10E+12 | 4.08E+05 +/- 1.44E+05 | 4.13E+05 | Takey+, 2011 |
| **J095603.5+410709** | 0.5708 | 2.93 | 4.31E+03 | 2.30E+13 | 5.30E+05 +/- 1.87E+05 | 4.65E+05 | Takey+, 2011 |
| **J095736.7+023427** | 0.3734 | 3.82 | 4.88E+03 | 2.28E+13 | 6.05E+05 +/- 2.14E+05 | 4.90E+05 | Takey+, 2011 |
| **J095902.7+025543** | 0.3479 | 2.22 | 6.56E+03 | 1.57E+13 | 4.61E+05 +/- 1.63E+05 | 4.67E+05 | Takey+, 2011 |
| **J095904.4+130524** | 0.3959 | 2.67 | 6.43E+03 | 4.13E+13 | 5.06E+05 +/- 1.79E+05 | 5.28E+05 | Takey+, 2011 |
| **J100021.9+022326** | 0.2213 | 0.99 | 6.41E+03 | 2.81E+12 | 3.08E+05 +/- 1.09E+05 | 3.68E+05 | Takey+, 2011 |
| **J100027.9+024129** | 0.3471 | 2.75 | 4.64E+03 | 5.89E+12 | 5.13E+05 +/- 1.81E+05 | 4.03E+05 | Takey+, 2011 |
| **J100117.4+285111** | 0.0946 | 0.83 | 5.02E+03 | 3.82E+12 | 2.82E+05 +/- 9.95E+04 | 3.95E+05 | Takey+, 2011 |
| **J100142.2+022543** | 0.1172 | 1.39 | 5.10E+03 | 4.39E+12 | 3.65E+05 +/- 1.29E+05 | 4.03E+05 | Takey+, 2011 |
| **J100200.9+020406** | 0.4617 | 1.14 | 6.01E+03 | 4.24E+12 | 3.30E+05 +/- 1.16E+05 | 3.74E+05 | Takey+, 2011 |
| **J100423.5+410008** | 0.3601 | 1.31 | 5.80E+03 | 4.01E+12 | 3.54E+05 +/- 1.25E+05 | 3.79E+05 | Takey+, 2011 |
| **J101042.4+554223** | 0.2597 | 1 | 4.61E+03 | 2.71E+12 | 3.09E+05 +/- 1.09E+05 | 3.63E+05 | Takey+, 2011 |
| **J101120.0+533431** | 0.3896 | 4.33 | 8.57E+03 | 1.12E+13 | 6.44E+05 +/- 2.27E+05 | 4.40E+05 | Takey+, 2011 |
| **J101930.3+080347** | 0.1801 | 2.93 | 5.70E+03 | 9.71E+12 | 5.30E+05 +/- 1.87E+05 | 4.51E+05 | Takey+, 2011 |
| **J102539.4+470502** | 0.1839 | 0.45 | 4.40E+03 | 2.95E+12 | 2.08E+05 +/- 7.34E+04 | 3.73E+05 | Takey+, 2011 |
| **J103007.0-030638** | 0.4274 | 2.21 | 6.25E+03 | 1.24E+13 | 4.60E+05 +/- 1.62E+05 | 4.43E+05 | Takey+, 2011 |
| **J104421.8+213029** | 0.4975 | 3.37 | 4.75E+03 | 1.35E+13 | 5.68E+05 +/- 2.00E+05 | 4.41E+05 | Takey+, 2011 |
| **J104423.0+064509** | 0.2989 | 2.36 | 5.46E+03 | 8.08E+12 | 4.75E+05 +/- 1.68E+05 | 4.28E+05 | Takey+, 2011 |
| **J104613.6+484820** | 0.4342 | 1.73 | 5.25E+03 | 8.31E+12 | 4.07E+05 +/- 1.44E+05 | 4.17E+05 | Takey+, 2011 |
| **J104619.4+483620** | 0.1409 | 1.45 | 7.56E+03 | 7.52E+12 | 3.73E+05 +/- 1.32E+05 | 4.37E+05 | Takey+, 2011 |
| **J105655.6+065840** | 0.296 | 2.32 | 8.06E+03 | 9.79E+12 | 4.71E+05 +/- 1.66E+05 | 4.41E+05 | Takey+, 2011 |
| **J110341.3+380512** | 0.3066 | 1.59 | 4.67E+03 | 5.19E+12 | 3.90E+05 +/- 1.38E+05 | 3.99E+05 | Takey+, 2011 |
| **J110344.0+360513** | 0.5356 | 1.44 | 6.81E+03 | 4.06E+12 | 3.71E+05 +/- 1.31E+05 | 3.65E+05 | Takey+, 2011 |
| **J111651.3+440945** | 0.5881 | 2.93 | 8.11E+03 | 6.30E+12 | 5.30E+05 +/- 1.87E+05 | 3.85E+05 | Takey+, 2011 |
| **J111712.5+423605** | 0.4404 | 2.44 | 5.20E+03 | 1.28E+13 | 4.83E+05 +/- 1.70E+05 | 4.43E+05 | Takey+, 2011 |
| **J111725.8+074337** | 0.4601 | 2.84 | 5.49E+03 | 1.23E+13 | 5.22E+05 +/- 1.84E+05 | 4.39E+05 | Takey+, 2011 |
| **J111729.9+174455** | 0.521 | 2.31 | 4.22E+03 | 1.32E+13 | 4.70E+05 +/- 1.66E+05 | 4.36E+05 | Takey+, 2011 |
| **J111803.5+403444** | 0.4207 | 1.16 | 4.82E+03 | 4.43E+12 | 3.33E+05 +/- 1.18E+05 | 3.80E+05 | Takey+, 2011 |
| **J111805.7+440935** | 0.3601 | 2.09 | 5.21E+03 | 5.22E+12 | 4.47E+05 +/- 1.58E+05 | 3.95E+05 | Takey+, 2011 |
| **J112259.2+465917** | 0.471 | 2.02 | 5.60E+03 | 9.12E+12 | 4.40E+05 +/- 1.55E+05 | 4.19E+05 | Takey+, 2011 |
| **J112419.9+470005** | 0.5337 | 1.83 | 4.22E+03 | 1.30E+13 | 4.19E+05 +/- 1.48E+05 | 4.34E+05 | Takey+, 2011 |
| **J112739.7+583720** | 0.1873 | 1.24 | 4.92E+03 | 4.55E+12 | 3.45E+05 +/- 1.22E+05 | 4.00E+05 | Takey+, 2011 |
| **J113843.7+031537** | 0.1196 | 1.89 | 5.94E+03 | 6.91E+12 | 4.25E+05 +/- 1.50E+05 | 4.33E+05 | Takey+, 2011 |
| **J113910.1+170344** | 0.297 | 1.53 | 6.81E+03 | 7.52E+12 | 3.83E+05 +/- 1.35E+05 | 4.23E+05 | Takey+, 2011 |
| **J114535.8+024250** | 0.1929 | 1.75 | 6.50E+03 | 4.74E+12 | 4.09E+05 +/- 1.44E+05 | 4.02E+05 | Takey+, 2011 |
| **J114541.4+025415** | 0.2779 | 0.79 | 5.18E+03 | 3.21E+12 | 2.75E+05 +/- 9.71E+04 | 3.71E+05 | Takey+, 2011 |
| **J114627.8+522108** | 0.453 | 1.56 | 6.58E+03 | 1.34E+13 | 3.87E+05 +/- 1.37E+05 | 4.45E+05 | Takey+, 2011 |
| **J115019.4+014054** | 0.4055 | 1.74 | 4.91E+03 | 4.27E+12 | 4.08E+05 +/- 1.44E+05 | 3.79E+05 | Takey+, 2011 |
| **J115040.7+545641** | 0.228 | 1.79 | 6.39E+03 | 2.25E+13 | 4.14E+05 +/- 1.46E+05 | 5.06E+05 | Takey+, 2011 |
| **J115125.3+545006** | 0.1463 | 1.83 | 4.36E+03 | 6.08E+12 | 4.19E+05 +/- 1.48E+05 | 4.22E+05 | Takey+, 2011 |
| **J115640.0+524317** | 0.5422 | 1 | 4.64E+03 | 4.92E+12 | 3.09E+05 +/- 1.09E+05 | 3.75E+05 | Takey+, 2011 |
| **J115818.8+440853** | 0.4321 | 1.54 | 4.79E+03 | 3.82E+12 | 3.84E+05 +/- 1.36E+05 | 3.70E+05 | Takey+, 2011 |
| **J115824.6+440532** | 0.4086 | 2.8 | 6.67E+03 | 1.07E+13 | 5.18E+05 +/- 1.83E+05 | 4.35E+05 | Takey+, 2011 |
| **J115850.5+440534** | 0.2901 | 0.82 | 4.50E+03 | 4.10E+12 | 2.80E+05 +/- 9.88E+04 | 3.85E+05 | Takey+, 2011 |
| **J120045.3+342454** | 0.2446 | 1.83 | 4.94E+03 | 4.13E+12 | 4.19E+05 +/- 1.48E+05 | 3.89E+05 | Takey+, 2011 |
| **J120124.3+342044** | 0.261 | 1.39 | 5.22E+03 | 5.06E+12 | 3.65E+05 +/- 1.29E+05 | 4.01E+05 | Takey+, 2011 |
| **J120130.9+341706** | 0.2748 | 1.99 | 7.46E+03 | 5.72E+12 | 4.37E+05 +/- 1.54E+05 | 4.08E+05 | Takey+, 2011 |
| **J120933.9+392234** | 0.5321 | 2.96 | 8.10E+03 | 8.24E+12 | 5.32E+05 +/- 1.88E+05 | 4.07E+05 | Takey+, 2011 |
| **J121205.3+131739** | 0.4489 | 1.73 | 6.67E+03 | 6.65E+12 | 4.07E+05 +/- 1.44E+05 | 4.02E+05 | Takey+, 2011 |
| **J121334.6+025347** | 0.4096 | 4.25 | 4.67E+03 | 1.89E+13 | 6.38E+05 +/- 2.25E+05 | 4.73E+05 | Takey+, 2011 |
| **J121744.1+472913** | 0.2707 | 2.94 | 9.14E+03 | 1.96E+13 | 5.31E+05 +/- 1.87E+05 | 4.91E+05 | Takey+, 2011 |
| **J121806.9+295847** | 0.3604 | 1.34 | 4.61E+03 | 4.21E+12 | 3.58E+05 +/- 1.26E+05 | 3.81E+05 | Takey+, 2011 |
| **J122527.7+004235** | 0.2369 | 1.95 | 6.43E+03 | 1.14E+13 | 4.32E+05 +/- 1.52E+05 | 4.57E+05 | Takey+, 2011 |
| **J122657.1+334332** | 0.4997 | 3.72 | 6.27E+03 | 2.59E+13 | 5.97E+05 +/- 2.11E+05 | 4.82E+05 | Takey+, 2011 |
| **J123040.1+115706** | 0.2514 | 1.14 | 6.26E+03 | 4.27E+12 | 3.30E+05 +/- 1.16E+05 | 3.91E+05 | Takey+, 2011 |
| **J123101.8+102855** | 0.1593 | 1.44 | 5.24E+03 | 5.66E+12 | 3.71E+05 +/- 1.31E+05 | 4.16E+05 | Takey+, 2011 |
| **J123145.0+413731** | 0.1718 | 2.3 | 5.24E+03 | 8.25E+12 | 4.69E+05 +/- 1.66E+05 | 4.41E+05 | Takey+, 2011 |
| **J123150.4+120005** | 0.2494 | 2.93 | 8.73E+03 | 1.58E+13 | 5.30E+05 +/- 1.87E+05 | 4.78E+05 | Takey+, 2011 |
| **J123601.3+263831** | 0.2098 | 1.36 | 9.36E+03 | 3.17E+12 | 3.61E+05 +/- 1.27E+05 | 3.76E+05 | Takey+, 2011 |
| **J123649.7+255002** | 0.1741 | 1.65 | 5.30E+03 | 4.73E+12 | 3.98E+05 +/- 1.40E+05 | 4.03E+05 | Takey+, 2011 |
| **J123841.0+092830** | 0.2299 | 4.93 | 5.08E+03 | 1.66E+13 | 6.87E+05 +/- 2.42E+05 | 4.84E+05 | Takey+, 2011 |
| **J124125.7-015824** | 0.1562 | 1.64 | 6.46E+03 | 5.49E+12 | 3.96E+05 +/- 1.40E+05 | 4.14E+05 | Takey+, 2011 |
| **J124202.5+332216** | 0.1367 | 0.66 | 5.77E+03 | 2.44E+12 | 2.51E+05 +/- 8.86E+04 | 3.64E+05 | Takey+, 2011 |
| **J124309.5+142026** | 0.3401 | 1.52 | 4.30E+03 | 1.89E+13 | 3.82E+05 +/- 1.35E+05 | 4.81E+05 | Takey+, 2011 |
| **J124425.3+164759** | 0.2346 | 1.1 | 7.77E+03 | 4.02E+12 | 3.25E+05 +/- 1.15E+05 | 3.88E+05 | Takey+, 2011 |
| **J124453.5-003335** | 0.2322 | 1.72 | 6.25E+03 | 5.07E+12 | 4.06E+05 +/- 1.43E+05 | 4.03E+05 | Takey+, 2011 |
| **J124600.4+543530** | 0.5327 | 1.81 | 5.46E+03 | 1.03E+13 | 4.16E+05 +/- 1.47E+05 | 4.20E+05 | Takey+, 2011 |
| **J124919.4+051837** | 0.3577 | 1.49 | 4.90E+03 | 4.19E+12 | 3.78E+05 +/- 1.33E+05 | 3.81E+05 | Takey+, 2011 |
| **J125003.9+052118** | 0.3531 | 1.04 | 5.50E+03 | 3.34E+12 | 3.16E+05 +/- 1.12E+05 | 3.68E+05 | Takey+, 2011 |
| **J125047.2+263338** | 0.4759 | 1.82 | 5.37E+03 | 8.33E+12 | 4.18E+05 +/- 1.48E+05 | 4.13E+05 | Takey+, 2011 |
| **J125540.1+255843** | 0.4637 | 1.09 | 7.73E+03 | 3.52E+12 | 3.23E+05 +/- 1.14E+05 | 3.63E+05 | Takey+, 2011 |
| **J125620.2+254603** | 0.2584 | 1.44 | 8.51E+03 | 4.39E+12 | 3.71E+05 +/- 1.31E+05 | 3.92E+05 | Takey+, 2011 |
| **J125752.4+283016** | 0.3585 | 2.08 | 5.87E+03 | 8.66E+12 | 4.46E+05 +/- 1.57E+05 | 4.27E+05 | Takey+, 2011 |
| **J130526.8+181200** | 0.4406 | 3.53 | 4.62E+03 | 1.43E+13 | 5.81E+05 +/- 2.05E+05 | 4.51E+05 | Takey+, 2011 |
| **J130534.9+175650** | 0.486 | 4.13 | 4.91E+03 | 1.32E+13 | 6.29E+05 +/- 2.22E+05 | 4.40E+05 | Takey+, 2011 |
| **J130554.1+181436** | 0.4416 | 1.36 | 5.76E+03 | 6.31E+12 | 3.61E+05 +/- 1.27E+05 | 3.99E+05 | Takey+, 2011 |
| **J130832.2+534215** | 0.329 | 1.83 | 8.63E+03 | 1.14E+13 | 4.19E+05 +/- 1.48E+05 | 4.47E+05 | Takey+, 2011 |
| **J130933.1+573928** | 0.2499 | 1.53 | 6.34E+03 | 4.51E+12 | 3.83E+05 +/- 1.35E+05 | 3.95E+05 | Takey+, 2011 |
| **J130947.7+573011** | 0.3151 | 1.8 | 4.49E+03 | 7.33E+12 | 4.15E+05 +/- 1.46E+05 | 4.20E+05 | Takey+, 2011 |
| **J131155.1+352342** | 0.1824 | 1.66 | 4.98E+03 | 4.03E+12 | 3.99E+05 +/- 1.41E+05 | 3.93E+05 | Takey+, 2011 |
| **J131303.2+351937** | 0.2991 | 2.91 | 5.71E+03 | 1.01E+13 | 5.28E+05 +/- 1.86E+05 | 4.42E+05 | Takey+, 2011 |
| **J132422.9+300930** | 0.6105 | 1.04 | 6.25E+03 | 4.59E+12 | 3.16E+05 +/- 1.12E+05 | 3.66E+05 | Takey+, 2011 |
| **J132631.5+074503** | 0.5758 | 1.05 | 5.03E+03 | 5.33E+12 | 3.17E+05 +/- 1.12E+05 | 3.77E+05 | Takey+, 2011 |
| **J132704.2+315129** | 0.236 | 1.26 | 4.40E+03 | 5.11E+12 | 3.47E+05 +/- 1.22E+05 | 4.04E+05 | Takey+, 2011 |
| **J132921.0-005358** | 0.55 | 2.32 | 6.69E+03 | 1.60E+13 | 4.71E+05 +/- 1.66E+05 | 4.45E+05 | Takey+, 2011 |
| **J133004.1+583429** | 0.3013 | 1.45 | 6.00E+03 | 4.26E+12 | 3.73E+05 +/- 1.32E+05 | 3.87E+05 | Takey+, 2011 |
| **J133434.6+375702** | 0.2696 | 1.34 | 5.04E+03 | 1.11E+13 | 3.58E+05 +/- 1.26E+05 | 4.52E+05 | Takey+, 2011 |
| **J133438.1+504331** | 0.2495 | 1.18 | 8.57E+03 | 3.33E+12 | 3.36E+05 +/- 1.19E+05 | 3.76E+05 | Takey+, 2011 |
| **J133458.4+375019** | 0.3996 | 2.09 | 6.35E+03 | 5.22E+12 | 4.47E+05 +/- 1.58E+05 | 3.91E+05 | Takey+, 2011 |
| **J133909.1+481149** | 0.4092 | 1.44 | 4.32E+03 | 6.24E+12 | 3.71E+05 +/- 1.31E+05 | 4.01E+05 | Takey+, 2011 |
| **J134138.7+001721** | 0.5081 | 2.93 | 5.05E+03 | 2.04E+13 | 5.30E+05 +/- 1.87E+05 | 4.66E+05 | Takey+, 2011 |
| **J134304.9-000053** | 0.6001 | 3 | 5.96E+03 | 2.92E+13 | 5.36E+05 +/- 1.89E+05 | 4.76E+05 | Takey+, 2011 |
| **J135358.9+334958** | 0.47 | 1.74 | 5.52E+03 | 1.39E+13 | 4.08E+05 +/- 1.44E+05 | 4.46E+05 | Takey+, 2011 |
| **J135540.9+182544** | 0.2585 | 1.11 | 7.05E+03 | 3.76E+12 | 3.26E+05 +/- 1.15E+05 | 3.83E+05 | Takey+, 2011 |
| **J135933.5+621901** | 0.3287 | 3.72 | 5.24E+03 | 3.05E+13 | 5.97E+05 +/- 2.11E+05 | 5.16E+05 | Takey+, 2011 |
| **J140612.4+282529** | 0.3441 | 0.62 | 5.24E+03 | 3.97E+12 | 2.44E+05 +/- 8.61E+04 | 3.79E+05 | Takey+, 2011 |
| **J140615.5+283051** | 0.5558 | 2.55 | 4.58E+03 | 1.39E+13 | 4.94E+05 +/- 1.74E+05 | 4.36E+05 | Takey+, 2011 |
| **J141457.7-002059** | 0.1427 | 1.32 | 6.36E+03 | 8.05E+12 | 3.56E+05 +/- 1.26E+05 | 4.41E+05 | Takey+, 2011 |
| **J141534.4+282342** | 0.2239 | 2.22 | 4.55E+03 | 8.73E+12 | 4.61E+05 +/- 1.63E+05 | 4.40E+05 | Takey+, 2011 |
| **J141553.9+281714** | 0.1388 | 1.51 | 4.40E+03 | 4.68E+12 | 3.80E+05 +/- 1.34E+05 | 4.05E+05 | Takey+, 2011 |
| **J141621.6+264310** | 0.457 | 1.04 | 7.93E+03 | 6.67E+12 | 3.16E+05 +/- 1.12E+05 | 4.01E+05 | Takey+, 2011 |
| **J141627.9+444645** | 0.3898 | 3.5 | 5.49E+03 | 2.84E+13 | 5.79E+05 +/- 2.04E+05 | 5.03E+05 | Takey+, 2011 |
| **J141830.6+251052** | 0.2909 | 5.92 | 9.86E+03 | 3.14E+13 | 7.53E+05 +/- 2.66E+05 | 5.23E+05 | Takey+, 2011 |
| **J142305.5+382807** | 0.4295 | 2 | 4.97E+03 | 6.69E+12 | 4.38E+05 +/- 1.55E+05 | 4.04E+05 | Takey+, 2011 |
| **J142731.1+262904** | 0.3416 | 1.5 | 6.40E+03 | 5.30E+12 | 3.79E+05 +/- 1.34E+05 | 3.97E+05 | Takey+, 2011 |
| **J143103.0+421432** | 0.4261 | 1.7 | 6.52E+03 | 1.20E+13 | 4.04E+05 +/- 1.43E+05 | 4.41E+05 | Takey+, 2011 |
| **J143121.1-005341** | 0.4029 | 3.48 | 5.94E+03 | 2.43E+13 | 5.77E+05 +/- 2.04E+05 | 4.90E+05 | Takey+, 2011 |
| **J143448.3+033749** | 0.1463 | 1.97 | 5.69E+03 | 6.29E+12 | 4.34E+05 +/- 1.53E+05 | 4.24E+05 | Takey+, 2011 |
| **J143649.3+632046** | 0.3333 | 0.75 | 6.14E+03 | 3.19E+12 | 2.68E+05 +/- 9.46E+04 | 3.67E+05 | Takey+, 2011 |
| **J143713.9+341519** | 0.5426 | 2.3 | 5.79E+03 | 1.25E+13 | 4.69E+05 +/- 1.66E+05 | 4.31E+05 | Takey+, 2011 |
| **J143850.1+341551** | 0.3407 | 1.13 | 6.36E+03 | 5.15E+12 | 3.29E+05 +/- 1.16E+05 | 3.95E+05 | Takey+, 2011 |
| **J143929.0+024605** | 0.1447 | 0.78 | 5.17E+03 | 2.79E+12 | 2.73E+05 +/- 9.64E+04 | 3.72E+05 | Takey+, 2011 |
| **J145014.9+270836** | 0.3294 | 1.83 | 4.40E+03 | 1.88E+13 | 4.19E+05 +/- 1.48E+05 | 4.82E+05 | Takey+, 2011 |
| **J145317.4+033446** | 0.3464 | 1.96 | 6.63E+03 | 9.92E+12 | 4.33E+05 +/- 1.53E+05 | 4.37E+05 | Takey+, 2011 |
| **J145709.1-010057** | 0.4764 | 1.42 | 4.58E+03 | 8.94E+12 | 3.69E+05 +/- 1.30E+05 | 4.17E+05 | Takey+, 2011 |
| **J145846.5+493501** | 0.4631 | 2.93 | 4.84E+03 | 1.39E+13 | 5.30E+05 +/- 1.87E+05 | 4.46E+05 | Takey+, 2011 |
| **J150824.5-001557** | 0.0889 | 1.41 | 5.15E+03 | 5.04E+12 | 3.68E+05 +/- 1.30E+05 | 4.14E+05 | Takey+, 2011 |
| **J151529.5+003943** | 0.2712 | 1.1 | 5.34E+03 | 6.00E+12 | 3.25E+05 +/- 1.15E+05 | 4.11E+05 | Takey+, 2011 |
| **J151747.9+424141** | 0.4151 | 1.23 | 4.89E+03 | 5.17E+12 | 3.43E+05 +/- 1.21E+05 | 3.89E+05 | Takey+, 2011 |
| **J153813.5+533318** | 0.4166 | 1.83 | 5.55E+03 | 9.54E+12 | 4.19E+05 +/- 1.48E+05 | 4.27E+05 | Takey+, 2011 |
| **J154406.5+534712** | 0.497 | 1.83 | 6.26E+03 | 8.05E+12 | 4.19E+05 +/- 1.48E+05 | 4.09E+05 | Takey+, 2011 |
| **J161040.5+540638** | 0.3618 | 1.07 | 7.56E+03 | 6.42E+12 | 3.20E+05 +/- 1.13E+05 | 4.07E+05 | Takey+, 2011 |
| **J161134.2+541635** | 0.3307 | 2.7 | 7.75E+03 | 1.85E+13 | 5.09E+05 +/- 1.80E+05 | 4.80E+05 | Takey+, 2011 |
| **J163249.1+401606** | 0.159 | 0.9 | 7.76E+03 | 2.81E+12 | 2.94E+05 +/- 1.04E+05 | 3.72E+05 | Takey+, 2011 |
| **J163431.7+330925** | 0.3321 | 1.54 | 5.88E+03 | 8.73E+12 | 3.84E+05 +/- 1.36E+05 | 4.30E+05 | Takey+, 2011 |
| **J170041.9+641258** | 0.235 | 3.7 | 4.66E+03 | 2.21E+13 | 5.95E+05 +/- 2.10E+05 | 5.04E+05 | Takey+, 2011 |
| **J213027.3-000017** | 0.1373 | 2.54 | 4.34E+03 | 6.55E+12 | 4.93E+05 +/- 1.74E+05 | 4.28E+05 | Takey+, 2011 |
| **MKW3S** | 0.045 | 3.44 | 5.22E+03 | 2.35E+13 | 5.74E+05 +/- 2.03E+05 | 5.24E+05 | Takey+, 2011 |
| **MKW8** | 0.027 | 2.5 | 6.77E+03 | 1.13E+13 | 4.89E+05 +/- 1.73E+05 | 4.71E+05 | Takey+, 2011 |
| **PKS0745-191** | 0.1028 | 6.76 | 8.24E+03 | 8.09E+13 | 8.05E+05 +/- 2.84E+05 | 6.15E+05 | Takey+, 2011 |
| **UGC3957** | 0.034 | 2.34 | 5.74E+03 | 1.05E+13 | 4.73E+05 +/- 1.67E+05 | 4.65E+05 | Takey+, 2011 |
| **ZwCl1215+0400** | 0.075 | 7.57 | 5.43E+03 | 6.57E+13 | 8.52E+05 +/- 3.01E+05 | 6.02E+05 | Takey+, 2011 |
| **ZwCl1742+3306** | 0.0757 | 4.46 | 4.39E+03 | 2.81E+13 | 6.54E+05 +/- 2.31E+05 | 5.35E+05 | Takey+, 2011 |
| **ZwIII54** | 0.0311 | 2.25 | 6.02E+03 | 1.14E+13 | 4.64E+05 +/- 1.64E+05 | 4.71E+05 | Takey+, 2011 |